\documentclass[a4paper,11pt,twocolumn]{article}

\usepackage{icphs2023}
\usepackage{metalogo} 
\usepackage{epstopdf}
\usepackage{amsmath}
\usepackage{amssymb}
\usepackage{tipa}

\title{Acoustic absement in detail: Quantifying acoustic differences across time-series representations of speech data}
\author{Matthew C. Kelley}
\organization{Department of Linguistics, University of Washington}
\email{mattck@uw.edu}
\begin{document}
	
	\maketitle
	
	\begin{abstract}
		The speech signal is a consummate example of time-series data. The acoustics of the signal change over time, sometimes dramatically. Yet, the most common type of comparison we perform in phonetics is between instantaneous acoustic measurements, such as formant values. In the present paper, I discuss the concept of absement as a quantification of differences between two time-series. I then provide an experimental example of absement applied to phonetic analysis for human and/or computer speech recognition. The experiment is a template-based speech recognition task, using dynamic time warping to compare the acoustics between recordings of isolated words. A recognition accuracy of 57.9\% was achieved. The results of the experiment are discussed in terms of using absement as a tool, as well as the implications of using acoustics-only models of spoken word recognition with the word as the smallest discrete linguistic unit.
	\end{abstract}
	
	\keywords{spoken word recognition, speech technology, acoustics, acoustic absement}
	
	\section{Introduction}
	
	The field of phonetics is replete with time-series and sequential data. Whether analyzing a waveform, spectrogram, or transcription, there is a natural and prescribed order to the sub-elements of the sequence---e.g., samples, spectra, or segments, respectively. Despite the sequential nature of so many of the types of data that we analyze, we often factor out time-bound aspects and instead focus on instantaneous measurements of a quantity, such as formant values or intensity at a given point in time. This choice is certainly convenient, and it is all but mandated by many of the statistical methods we employ in speech science.
	
	Nevertheless, the field stands to benefit from incorporating more methods of analysis that can account for time-series data. In the present paper, I focus on the quantity of absement as a concrete example of a measurement that reflects the temporal nature of speech data. In the present paper, I present a simplified speech recognition experiment using dynamic time warping. This task itself is a classical example of early speech recognition techniques, and the purpose is to demonstrate a use for absement for phonetic analysis. However, it also bears some relation to models of spoken word recognition, which I will also discuss.
	
	\subsection{Theoretical background}
	
	Before formally introducing absement, it is necessary to discuss distance. Distance itself is a familiar notion, often quantifying how far apart two objects are in physical space. An analogous example from phonetics would be how far apart two formant values are from each other. There are infinite methods to calculate distance, but in mechanics, distance is defined as the magnitude of a displacement vector, which is the same as the formula for Euclidean distance. Euclidean distance itself is well-attested in phonetics, such as when comparing the centroids of an F1-by-F2 space for two vowel categories \cite{kendallVariationPerceptionProduction2012}. Formally, the Euclidean distance $d$ between two vectors $x$ and $y$ of length $k$ is
	
	\begin{equation}
		\label{eq:norm}
		d(x, y) = \Vert x - y \Vert_2 = \sqrt{\sum_{i=1}^k \vert \chi_i - \psi_i \vert^2}\,,
	\end{equation}
	
	\noindent where $\chi_i$ and $\psi_i$ are the scalar elements within $x$ and $y$ at index $i$, respectively.
	
	If the distance between two objects is summed or integrated over time, the result is a quantity referred to in \cite{mannHydraulophoneDesignConsiderations2006} and \cite{mannEffectivenessIntegralKinesiology2018} as ``absement'' [\textprimstress\ae\textipa{bs@m\s{n}t}], a blend of \textit{absence} and \textit{displacement}. This is the time-distance product form of absement, though a time-displacement product form also exists \cite{mannHydraulophoneDesignConsiderations2006}. Absement is a measure of prolonged distance. For time-series data, absement can be thought of as how well two objects matched each other over time. A high absement value could be caused by two objects being apart over a long period of time, a moment of extreme distance, or a combination thereof. Formally, the absement between two time-series $X$ and $Y$ is
	
	\begin{equation}
		a(X, Y) = \int_0^T d(x_t, y_t) \, \mathrm{d}t\,,
	\end{equation}
	
	\noindent where $x_t$ and $y_t$ are vectors representing the value of $X$ and $Y$ at time $t$, and $T$ is the total length of time over which absement is being calculated.
	
	Absement naturally accounts for the acoustic events that happen over a span of time, in a way that an instantaneous measure cannot. We do employ methods in phonetics that work well with time-series like the Fourier transform. But, these methods are often used to obtain an instantaneous measurement like a formant or fundamental frequency, rather than as the actual object of analysis. Tools like absement provide an opportunity to incorporate more of the time-varying nature of speech into our formal analyses.
	
	\subsection{Dynamic time warping as absement}
	
\begin{figure}
		\centering
		\includegraphics[width=0.9\columnwidth]{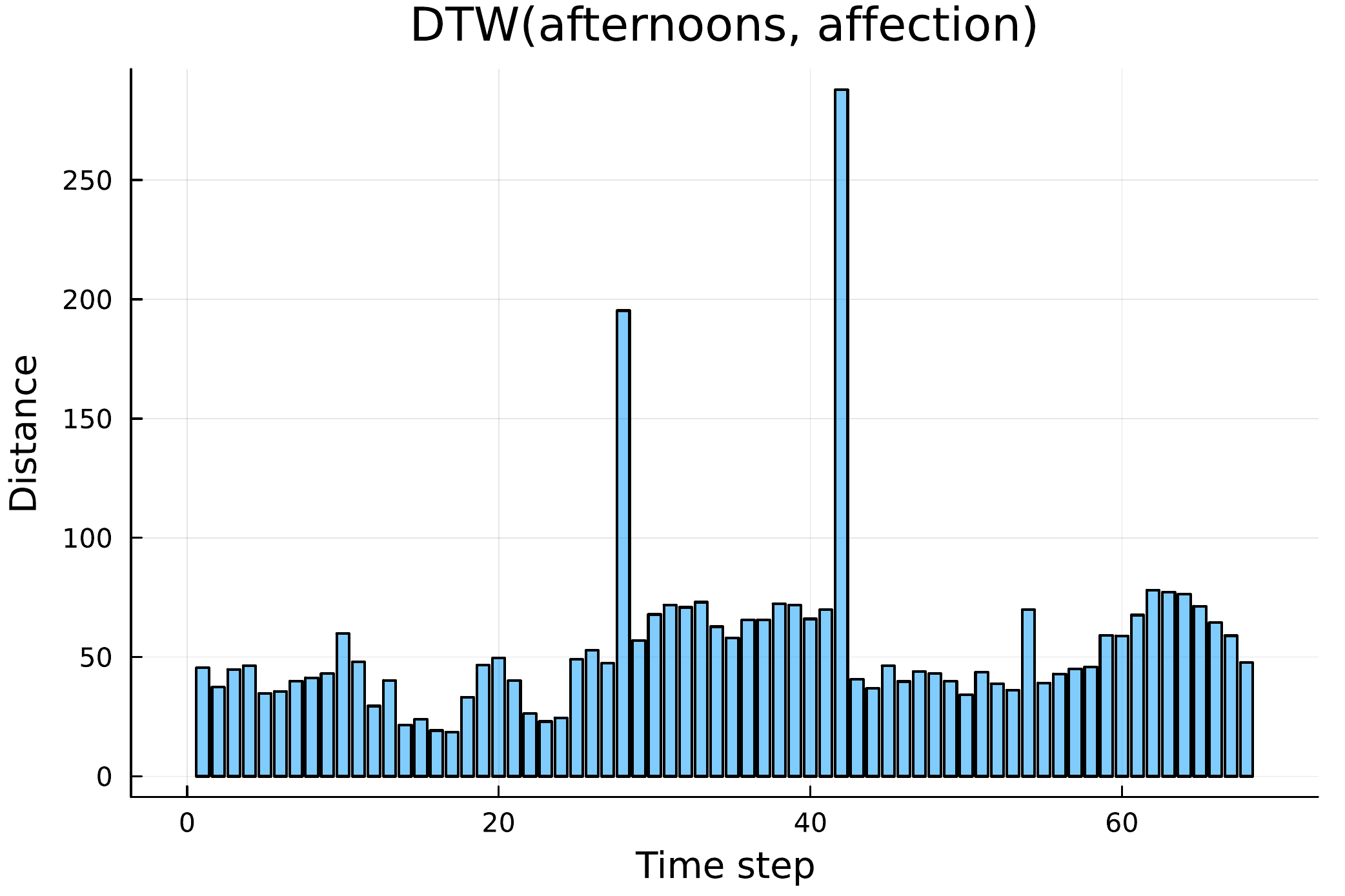}
		\caption{Distance values over time calculated from dynamic time warping between \textit{afternoons} and \textit{affection}. When a time point in \textit{afternoons} was stretched over multiple time points in \textit{affection}, those values were summed. Summing the area contained in each column will yield the dynamic time warping cost qua absement.}
		\label{fig:distance}
	\end{figure}

	Perhaps the most common calculation of absement in phonetics has been dynamic time warping. Dynamic time warping is a function that compares two time-series and yields an overall difference or cost value. When the comparison is performed, distance is calculated between each pairwise combination of time steps in the two signals. Then, dynamic programming is used to find a nonlinear warping path indicating which time steps should be compared between the two signals to minimize the overall computed difference between them. This discrete sum over time is roughly analogous to a rectangle method of approximation to the Riemann integral of distance. An example of the distance values calculated over time between \textit{afternoons} and \textit{affection} with dynamic time warping is shown in Figure~\ref{fig:distance}.
	
	The nonlinear warping path is crucial for several reasons. First, it allows for time series of different lengths to be compared, and the durations of any two productions are rarely the same. Second, it allows elasticity in the comparisons, which is important since word and segment durations vary between different productions in speech without resulting in wildly different percepts. Finally, this process allows for similar regions of each signal to be compared, permitting, for example, vowels in one signal to be compared to vowels in the other, and not necessarily to consonants.
	
	The shortcoming, however, is that the nonlinear warping path somewhat perturbs the natural analogy to absement from mechanics. While dynamic time warping does sum distance over time, the movement through time is not linear or uniform, and some moments in time would have multiple distance measurements associated with them. Regardless, I maintain that it is useful to refer to the output of dynamic time warping as absement to differentiate it from distance between two instantaneous measurements.
	
	Dynamic time warping as absement has found recent explicit use in spoken word recognition \cite{kelleyUsingAcousticDistance2022}, and previous research has also employed dynamic time warping without invoking absement \cite{bennettStatisticalAcousticEffects2018}. Dynamic time warping dates back to the 1970s \cite{sakoe1970similarity, sakoeDynamicProgrammingAlgorithm1978}, when it was developed to serve as the computational engine undergirding template-based automatic speech recognition. The field of automatic speech recognition has by and large moved away from dynamic time warping, first to hidden Markov models paired with Gaussian mixtures, of which dynamic time warping is a special case \cite{juangHiddenMarkovModel1984}. Modern speech recognition makes prevalent use of deep neural network models. Yet dynamic time warping has persisted as a useful method in data science \cite{rakthanmanonSearchingMiningTrillions2012,wuFastDTWApproximateGenerally2022} and cognitive science \cite{schatzEarlyPhoneticLearning2021,milletSelfsupervisedSpeechModels2022}.
	
	\section{Methods}
	
	The speech recognition experiment takes the form of an isolated word recognition task. The task is performed over a lexicon of 1,000 words and serves to demonstrate how absement can be a useful concept in phonetics.
	
	\subsection{Materials}
	
	The audio data came from the Massive Auditory Lexical Decision (MALD) project from Tucker et al. \cite{tuckerMassiveAuditoryLexical2019}. The project comprises responses to over 26,000 different English words in an auditory lexical decision task. The stimuli for the initial experiment were from a young male speaker of Canadian English who was trained in phonetics. Two other speakers were also recorded for auditory stimuli. These speakers were a young female speaker and an older male speaker of Canadian English. More recording information is given in \cite{tuckerMassiveAuditoryLexical2019}.
	
	I randomly sampled 1,000 of the words from the project that were recorded by all speakers. I converted each recording to a mel frequency cepstral coefficient (MFCC)-by-time representation using \texttt{MFCC.jl} v0.3.3 \cite{vanleeuwenMFCCJl2022} in \texttt{Julia} v1.8.2 \cite{bezansonJuliaFreshApproach2017}. The window length was 25 ms with an advance of 10 ms. 13 coefficients were calculated, and the first coefficient was replaced with log energy, as is standard in automatic speech recognition.
	
	I then used dynamic barycenter averaging \cite{petitjeanGlobalAveragingMethod2011} to create an average between the young female speaker and the older male speaker's recordings for each word. The averaging was performed using the phonetic sequence averaging interface in \texttt{Phonetics.jl} v0.1.2 \cite{kelleyPhoneticsJl2022}, which relies on \texttt{DynamicAxisWarping.jl} v0.4.12 \cite{baggecarlsonDynamicAxisWarpingJl2022}. In the averaging process, I randomly selected which of the two recordings of each word would serve as the basic template for the initial average sequence in the dynamic barycenter averaging process.
	
	The MFCC representation of the young male speaker was used to simulate a listener hearing someone speaking. The averages between the younger female and older male speakers' word productions were used as a set of acoustic templates that the listener would discriminate between based on the incoming acoustic signal. That is, the averaged productions were used as an acoustic representation of a small lexicon. I note that ``representation'' here is used in a general sense of one thing that stands in for another, and not necessarily as a cognitive model of language. Dynamic time warping in this sense can be approximately thought of as quantifying how different the spectrograms are between the young male speaker's productions and the spectrograms averaged between the young female and older male speakers' productions.
	
	\subsection{Analysis}
	
	Absement between each of the young male speaker's recordings and each of the average recordings was calculated using dynamic time warping with \texttt{DynamicAxisWarping.jl}. The distance function was Euclidean distance, and no warping radius was used. It is important to recall that absement can be increased both by duration and distance. While it is reasonable for duration to affect the absement value to some degree in spoken word recognition, the dynamic time warping calculation is such that shorter words tend to have smaller absement values on average, which biases recognition to short words.
	
	To overcome this undesirable behavior, some of the length discrepancy must be factored out. However, care must be used when performing this kind of factorization because simply dividing by the length in time transforms the absement value into average distance, which is no longer absement \textit{per se}. This relationship is analogous to how dividing distance by time yields average speed. A manual search of scaling functions suggested that dividing the absement values by the square root of the length of the averaged template helped normalize away some of the undesirable effect of duration on the absement values without completely destroying the ability to interpret the values as absement.
	
	The scaled absement values for each word were then sorted, and the top ten words in the lexicon with the lowest absement values for a given word were recorded. The word with the lowest absement value was taken as the word that was ultimately recognized given the acoustic input.
	
	\section{Results and discussion}
	
	Of the 1,000 words compared, 57.9\% were identified correctly based on the scaled absement values, and 87.9\% were in the top ten. These results compare favorably to some recent models of spoken word recognition using naive discriminative learning. Arnold et al. \cite{arnoldWordsSpontaneousConversational2017} reported an accuracy of 25.2\% on a 1,000 word recognition task, and Shafei-Bajestan and Baayen \cite{shafaei-bajestanWideLearningAuditory2018} reported an average best accuracy of 11.72\% correct on their various tasks for clean speech.
	
	The higher recognition accuracy in the present data must be considered with some caveats. The main limitation is that a 1,000 word lexicon is too limited to be a good reflection of a proficient speaker's knowledge. Additionally, as more words are introduced into the lexicon, the accuracy will almost assuredly fall. Discrimination tasks generally become harder when there are more items to discriminate between, due to the curse of dimensionality. Indeed, both naive discriminative learning models used a much fuller set of over 10,000 lexomes in their lexicons, regardless of task. Their lexicons were also based on corpora of connected conversational speech, so they were ultimately performing a harder task than the one I performed here.
	
	Regardless of any potential inflation in the accuracy I reported, I want to highlight that it would be much more difficult to recognize words using comparisons of instantaneous measurements. An acoustic measurement taken at a specific point in time is an infinitesimal portion of the acoustic signal and is insufficient to account for the communicative purpose of speech. Previous research has highlighted the need to situate phonetic measurements and cues in the communicative system they are used in \cite{redmonInterfacesSystemEmbedding2022,redmonLexicalAcousticsLinking2020}, discriminating between different words (or forms) and their associated meanings as in \cite{baayenDiscriminativeLexiconUnified2019}. Incorporating tools like absement into our methodologies is one way to better account for the temporal structure of the speech signal.
	
	It is also important to note that the model I have presented in the present paper does not make use of any sublexical unit like a segment, phone, or phoneme. The only possible sublexical units in the method I used are subsets of the MFCC matrices that represent the words in the lexicon. In the context of the present paper, I do not wish to interpret this as either evidence for or against the ontological or psychological reality of traditional sublexical units in linguistics. However, some previous models of spoken word recognition have proposed eschewing these types of units in favor of wholesale spoken word discrimination based solely on acoustic information \cite{arnoldWordsSpontaneousConversational2017,shafaei-bajestanWideLearningAuditory2018,baayenDiscriminativeLexiconUnified2019,shafaei-bajestanLDLAURISComputationalModel2021}.
	
	I do think that it will be worthwhile to explore this type of spoken word recognition model further. But, we should also keep in mind that this type of word recognition directly from acoustic data hails from a venerable idea from the early days of modern speech technology. It has also been reiterated at various points throughout the field's history. Exemplar theory models based on whole-word acoustics are one example \cite{johnsonAuditoryPerceptualBasis1997}. Goldinger \cite{goldingerPuzzlesolvingScienceQuixotic2003} also argued that many linguistic units measured in speech perception tasks emerge based on experimental task. Making explicit use of absement between acoustic representations may help push these lines of thinking forward to either present more convincing accounts of or eliminate certain hypotheses about spoken word recognition.
	
	Absement is a general concept and can be applied across a wide variety of tasks. It could, for example, be used to quantify the differences between vowel formant trajectories. It has indeed already seen similar such use via dynamic time warping on pitch contours \cite{mccloyProsodyIntelligibilityFamiliarity2013}. Virtually any comparison between a series of acoustic measurements made over time is amenable to being treated with absement.
	
	There is also more to be said about the use of dynamic time warping to instantiate and measure acoustic absement between two words. Because dynamic time warping has a propensity to allow duration to have an outsized effect on the output value, and because it can have certain time points repeat in the warping path, it is an imperfect representation of the concept of absement. The resolution I used here of scaling the absement values for better recognition accuracy is tantamount to having a constant weight function within the integral of distance. It may be worth exploring more sophisticated weight functions in the future or finding algorithms for calculating absement that do not require weighting.
	
	The code used to perform the analysis and a table of the results is available as supplementary material at \url{https://doi.org/10.5281/zenodo.7823844} or on GitHub at \url{https://github.com/maetshju/absement_in_detail_code}.
	
	\section{Conclusion}
	
	Absement is not an entirely new concept in phonetics. However, having a label for the type of time-series comparisons I have exhibited in the present paper should make working with this concept easier. Such a label may also nudge the idea of time-series analysis closer to the fore of our minds. Although, I am sure that none among us in phonetics truly need reminding of the time-varying nature of speech.
	
	\bibliographystyle{IEEEtran}
	\bibliography{references}

\begin{thebibliography}{10}
\providecommand{\url}[1]{#1}
\csname url@samestyle\endcsname
\providecommand{\newblock}{\relax}
\providecommand{\bibinfo}[2]{#2}
\providecommand{\BIBentrySTDinterwordspacing}{\spaceskip=0pt\relax}
\providecommand{\BIBentryALTinterwordstretchfactor}{4}
\providecommand{\BIBentryALTinterwordspacing}{\spaceskip=\fontdimen2\font plus
\BIBentryALTinterwordstretchfactor\fontdimen3\font minus
  \fontdimen4\font\relax}
\providecommand{\BIBforeignlanguage}[2]{{%
\expandafter\ifx\csname l@#1\endcsname\relax
\typeout{** WARNING: IEEEtran.bst: No hyphenation pattern has been}%
\typeout{** loaded for the language `#1'. Using the pattern for}%
\typeout{** the default language instead.}%
\else
\language=\csname l@#1\endcsname
\fi
#2}}
\providecommand{\BIBdecl}{\relax}
\BIBdecl

\bibitem{kendallVariationPerceptionProduction2012}
T.~Kendall and V.~Fridland, ``Variation in perception and production of mid
  front vowels in the {{U}}.{{S}}. {{Southern Vowel Shift}},'' \emph{Journal of
  Phonetics}, vol.~40, no.~2, pp. 289--306, Mar. 2012.

\bibitem{mannHydraulophoneDesignConsiderations2006}
S.~Mann, R.~Janzen, and M.~Post, ``Hydraulophone design considerations:
  Absement, displacement, and velocity-sensitive music keyboard in which each
  key is a water jet,'' in \emph{Proceedings of the 14th {{ACM}} International
  Conference on {{Multimedia}}}, ser. {{MM}} '06.\hskip 1em plus 0.5em minus
  0.4em\relax {New York, NY, USA}: {Association for Computing Machinery}, Oct.
  2006, pp. 519--528.

\bibitem{mannEffectivenessIntegralKinesiology2018}
S.~Mann, M.~L. Hao, M.~Tsai, M.~Hafezi, A.~Azad, and F.~Keramatimoezabad,
  ``Effectiveness of {{Integral Kinesiology Feedback}} for {{Fitness-Based
  Games}},'' in \emph{2018 {{IEEE Games}}, {{Entertainment}}, {{Media
  Conference}} ({{GEM}})}, Aug. 2018, pp. 1--9.

\bibitem{kelleyUsingAcousticDistance2022}
M.~C. Kelley and B.~V. Tucker, ``Using acoustic distance and acoustic absement
  to quantify lexical competition,'' \emph{The Journal of the Acoustical
  Society of America}, vol. 151, no.~2, pp. 1367--1379, Feb. 2022.

\bibitem{bennettStatisticalAcousticEffects2018}
R.~Bennett, K.~Tang, and J.~A. Sian, ``{Statistical and acoustic effects on the
  perception of stop consonants in Kaqchikel (Mayan)},'' \emph{Laboratory
  Phonology: Journal of the Association for Laboratory Phonology}, vol.~9,
  no.~1, p.~9, May 2018.

\bibitem{sakoe1970similarity}
H.~Sakoe and S.~Chiba, ``A similarity evaluation of speech patterns by dynamic
  programming,'' in \emph{Nat. {{Meeting}} of Institute of Electronic
  Communications Engineers of Japan}, 1970, p. 136.

\bibitem{sakoeDynamicProgrammingAlgorithm1978}
------, ``Dynamic programming algorithm optimization for spoken word
  recognition,'' \emph{IEEE Transactions on Acoustics, Speech, and Signal
  Processing}, vol.~26, no.~1, pp. 43--49, Feb. 1978.

\bibitem{juangHiddenMarkovModel1984}
B.-H. Juang, ``On the {{Hidden Markov Model}} and {{Dynamic Time Warping}} for
  {{Speech Recognition}}\textemdash{{A Unified View}},'' \emph{AT\&T Bell
  Laboratories Technical Journal}, vol.~63, no.~7, pp. 1213--1243, 1984.

\bibitem{rakthanmanonSearchingMiningTrillions2012}
T.~Rakthanmanon, B.~Campana, A.~Mueen, G.~Batista, B.~Westover, Q.~Zhu,
  J.~Zakaria, and E.~Keogh, ``Searching and mining trillions of time series
  subsequences under dynamic time warping,'' in \emph{Proceedings of the 18th
  {{ACM SIGKDD}} International Conference on {{Knowledge}} Discovery and Data
  Mining}, ser. {{KDD}} '12.\hskip 1em plus 0.5em minus 0.4em\relax {New York,
  NY, USA}: {Association for Computing Machinery}, Aug. 2012, pp. 262--270.

\bibitem{wuFastDTWApproximateGenerally2022}
R.~Wu and E.~J. Keogh, ``{{FastDTW}} is {{Approximate}} and {{Generally Slower
  Than}} the {{Algorithm}} it {{Approximates}},'' \emph{IEEE Transactions on
  Knowledge and Data Engineering}, vol.~34, no.~8, pp. 3779--3785, Aug. 2022.

\bibitem{schatzEarlyPhoneticLearning2021}
T.~Schatz, N.~H. Feldman, S.~Goldwater, X.-N. Cao, and E.~Dupoux, ``Early
  phonetic learning without phonetic categories: {{Insights}} from large-scale
  simulations on realistic input,'' \emph{Proceedings of the National Academy
  of Sciences}, vol. 118, no.~7, p. e2001844118, Feb. 2021.

\bibitem{milletSelfsupervisedSpeechModels2022}
J.~Millet and E.~Dunbar, ``Do self-supervised speech models develop human-like
  perception biases?'' in \emph{Proceedings of the 60th {{Annual Meeting}} of
  the {{Association}} for {{Computational Linguistics}} ({{Volume}} 1: {{Long
  Papers}})}.\hskip 1em plus 0.5em minus 0.4em\relax {Dublin, Ireland}:
  {Association for Computational Linguistics}, May 2022, pp. 7591--7605.

\bibitem{tuckerMassiveAuditoryLexical2019}
B.~V. Tucker, D.~Brenner, D.~K. Danielson, M.~C. Kelley, F.~Nenadi{\'c}, and
  M.~Sims, ``The {{Massive Auditory Lexical Decision}} ({{MALD}}) database,''
  \emph{Behavior Research Methods}, vol.~51, no.~3, pp. 1187--1204, Jun. 2019.

\bibitem{vanleeuwenMFCCJl2022}
\BIBentryALTinterwordspacing
D.~{van Leeuwen}, ``{{MFCC}}.jl,'' 2022. [Online]. Available:
  \url{https://github.com/JuliaDSP/MFCC.jl}
\BIBentrySTDinterwordspacing

\bibitem{bezansonJuliaFreshApproach2017}
J.~Bezanson, A.~Edelman, S.~Karpinski, and V.~Shah, ``Julia: {{A Fresh
  Approach}} to {{Numerical Computing}},'' \emph{SIAM Review}, vol.~59, no.~1,
  pp. 65--98, Jan. 2017.

\bibitem{petitjeanGlobalAveragingMethod2011}
F.~Petitjean, A.~Ketterlin, and P.~Gan{\c c}arski, ``A global averaging method
  for dynamic time warping, with applications to clustering,'' \emph{Pattern
  Recognition}, vol.~44, no.~3, pp. 678--693, Mar. 2011.

\bibitem{kelleyPhoneticsJl2022}
\BIBentryALTinterwordspacing
M.~C. Kelley, ``Phonetics.jl,'' 2022. [Online]. Available:
  \url{https://github.com/maetshju/Phonetics.jl}
\BIBentrySTDinterwordspacing

\bibitem{baggecarlsonDynamicAxisWarpingJl2022}
\BIBentryALTinterwordspacing
F.~Bagge~Carlson, ``{{DynamicAxisWarping}}.jl,'' 2022. [Online]. Available:
  \url{https://github.com/baggepinnen/DynamicAxisWarping.jl}
\BIBentrySTDinterwordspacing

\bibitem{arnoldWordsSpontaneousConversational2017}
D.~Arnold, F.~Tomaschek, K.~Sering, F.~Lopez, and R.~H. Baayen, ``Words from
  spontaneous conversational speech can be recognized with human-like accuracy
  by an error-driven learning algorithm that discriminates between meanings
  straight from smart acoustic features, bypassing the phoneme as recognition
  unit,'' \emph{PLOS ONE}, vol.~12, no.~4, p. e0174623, Apr. 2017.

\bibitem{shafaei-bajestanWideLearningAuditory2018}
E.~{Shafaei-Bajestan} and R.~H. Baayen, ``Wide learning for auditory
  comprehension,'' in \emph{Interspeech 2018}.\hskip 1em plus 0.5em minus
  0.4em\relax {ISCA}, Sep. 2018, pp. 966--970.

\bibitem{redmonInterfacesSystemEmbedding2022}
C.~Redmon and A.~Jongman, ``From interfaces to system embedding: {{Phonetic}}
  contrasts in the lexicon,'' Nov. 2022.

\bibitem{redmonLexicalAcousticsLinking2020}
C.~H. Redmon, ``Lexical acoustics: {{Linking}} phonetic systems to the
  higher-order units they encode,'' Ph.D. dissertation, University of Kansas,
  2020.

\bibitem{baayenDiscriminativeLexiconUnified2019}
R.~H. Baayen, Y.-Y. Chuang, E.~{Shafaei-Bajestan}, and J.~P. Blevins, ``The
  discriminative lexicon: {{A}} unified computational model for the lexicon and
  lexical processing in comprehension and production grounded not in
  (de)composition but in linear discriminative learning,'' \emph{Complexity},
  vol. 2019, 2019.

\bibitem{shafaei-bajestanLDLAURISComputationalModel2021}
E.~{Shafaei-Bajestan}, M.~{Moradipour-Tari}, P.~Uhrig, and R.~H. Baayen,
  ``{{LDL-AURIS}}: {{A}} computational model, grounded in error-driven
  learning, for the comprehension of single spoken words,'' \emph{Language,
  Cognition and Neuroscience}, vol.~0, no.~0, pp. 1--28, Jul. 2021.

\bibitem{johnsonAuditoryPerceptualBasis1997}
K.~Johnson, ``The auditory/perceptual basis for speech segmentation,'' {Ohio
  State University Department of Linguistics}, Working {{Paper}}, Jul. 1997.

\bibitem{goldingerPuzzlesolvingScienceQuixotic2003}
S.~D. Goldinger and T.~Azuma, ``Puzzle-solving science: The {{Quixotic}} quest
  for units in speech perception,'' \emph{Journal of Phonetics}, vol.~31,
  no.~3, pp. 305--320, Jul. 2003.

\bibitem{mccloyProsodyIntelligibilityFamiliarity2013}
D.~R. McCloy, ``Prosody, intelligibility and familiarity in speech
  perception,'' Ph.D. dissertation, University of Washington, Jul. 2013.

\end{thebibliography}
	
\end{document}